\documentclass[preprint,12pt]{elsarticle}

\usepackage{amssymb}
\usepackage{graphicx}
\usepackage{amsmath}
\usepackage{multirow}
\usepackage{makecell}
\usepackage{subfigure}
\usepackage{float}
\usepackage{algorithm}
\usepackage{algorithmic}
\usepackage{hyperref}
\usepackage[misc]{ifsym}

\journal{AAA}

\begin{document}

\begin{frontmatter}



\title{Utilizing Citation Network Structure to Predict Citation Counts: A Deep Learning Approach}


\author{Qihang Zhao\corref{cor1}}
\cortext[cor1]{zhaoqh75@mail.ustc.edu.cn}
\address{School of Management, University of Science and Technology of China, Hefei, Anhui Province, China}

\begin{abstract}
With the advancement of science and technology, the number of academic papers published in the world each year has increased almost exponentially. While a large number of research papers highlight the prosperity of science and technology, they also give rise to some problems. As we all know, academic papers are the most intuitive embodiment of the research results of scholars, which can reflect the level of researchers. It is also the evaluation standard for decision-making such as promotion and allocation of funds. Therefore, how to measure the quality of an academic paper is very important. The most common standard for measuring academic papers is the number of citation counts of papers, because this indicator is widely used in the evaluation of scientific publications, and it also serves as the basis for many other indicators (such as the h-index). Therefore, it is very important to be able to accurately predict the citation counts of academic papers.

This paper proposes an end-to-end deep learning network, DeepCCP, which combines the effect of information cascade and looks at the citation counts prediction problem from the perspective of information cascade prediction. DeepCCP directly uses the citation network formed in the early stage of the paper as the input, and the output is the citation counts of the corresponding paper after a period of time. DeepCCP only uses the structure and temporal information of the citation network, and does not require other additional information, but it can still achieve outstanding performance. According to experiments on 6 real data sets, DeepCCP is superior to the state-of-the-art methods in terms of the accuracy of citation count prediction.
\end{abstract}




\end{frontmatter}


\section{Introduction}
\label{sec_intro}
Nowadays, with the development of science and technology, many researchers are engaged in scientific research projects and writing research papers on a global scale. As a result, a large number of academic papers are published in academic journals or conferences every day, and they are increasing exponentially every year. As we all know, academic papers are the intuitive embodiment of researchers' research results. They not only play an important role in scientific evaluation such as recruitment decision-making, promotion, allocation of funds(\cite{bai2019predicting}, \cite{li2015trend}, \cite{chan2018relation}, \cite{fiala2017pagerank}, \cite{havemann2015bibliometric}), but also an important reference factor to measure the level of a researcher. Therefore, how to measure the quality of academic papers is crucial. Among the several existing quality standards for evaluating papers, the number of paper citations is undoubtedly the most important indicator(\cite{didegah2013factors}, \cite{garfield1998use}, \cite{moed2006citation}, \cite{oppenheim1995correlation}, \cite{yan2012better}), because it is widely used to measure the impact of papers. In addition, the citations of papers are also used as the basis for many other evaluation indicators, such as h-index(\cite{hirsch2005index}), journal impact factor(\cite{garfield2006history}), h5-index(\cite{zhang2017evaluating}), i-10 index and others Evaluation indicators for academic conferences, research institutions, etc. Therefore, predicting the counts of citation of a paper is very valuable and meaningful.

Predicting the citations of papers is a very complicated problem, and many scholars classify it as regression or classification problem(\cite{ruan2020predicting}). For the classification problem, some scholars simply use SVM, k-NN (\cite{fu2008models}, \cite{fu2010using}, \cite{wang2019can}) and other machine learning methods to predict whether the citation counts of the paper is higher than a specific value in the subsequent time, and divides the papers into two categories or several categories(\cite{ibanez2009predicting}, \cite{wang2011mining}) (such as high-level, medium level, low-level, etc.);
It is a more common method to classify the prediction of paper citation counts as regression problem (\cite{lokker2008prediction}, \cite{yu2014citation}). This kind of method extracts the relevant features of the paper, such as the characteristics of the authors(\cite{aksnes2003characteristics}), the time feature(\cite{xiao2016modeling}, \cite{abrishami2019predicting}), the content feature(\cite{wen2020paper}), etc, and then uses machine learning or stochastic process(\cite{xiao2016modeling}, \cite{lee2007depth}) to predict the specific citation counts of the paper, and achieves good results. However, the problem is that the citation patterns of papers are diverse, such as the phenomenon of "Sleeping Beauty in Science"(\cite{ke2015defining}, \cite{van2004sleeping}). Therefore, the traditional single and simple method may be powerless in the face of such diverse citation patterns, so a more powerful method is urgently needed.

Recently, with the continuous development of deep learning technology, scholars have begun to use deep learning related technologies in various fields, such as natural language processing(\cite{severyn2015learning}, \cite{sutskever2011generating}), speech recognition(\cite{hinton2012deep}, \cite{dahl2011context}), image processing(\cite{he2016deep}, \cite{wan2014deep}), network embedding(\cite{grover2016node2vec}, \cite{perozzi2014deepwalk}) and so on, and great progress has been made in these areas. Some scholars were inspired by this and began to introduce deep learning related methods in the field of citation counts prediction(\cite{ruan2020predicting}, \cite{abrishami2019predicting}, \cite{wen2020paper}, \cite{acuna2012predicting}). However, although these methods have achieved relatively good results in the citation counts prediction scenario, these methods still need to manually extract the features of the paper or only adopt a relatively shallow deep learning model, so they cannot effectively extract the citation information.

In order to solve such shortcomings, we developed an end-to-end deep learning model for paper citation prediction: DeepCCP, which combines the information cascade effect to compare paper citation counts prediction from the perspective of information cascade prediction. DeepCCP mainly takes the cascade citation network formed by citations in early stage of papers(such as 3-5 years) as input. The output of the model is the increase in the scale of this citation network over a period of time, that is, the amount of a certain paper in the future. The complex deep learning network structure of DeepCCP could capture the deep structure information of the citation network and the temporal information of the paper citation generation, and then can achieve the state-of-the-art in the paper citation counts prediction scenario. DeepCCP does not require other information, such as author characteristics, journal attributes or paper content, and it dynamically models paper citation networks, so different citation forms can be applied, that is, DeepCCP has a more powerful universality. We also conducted comprehensive experiments on 2 types of 6 real data sets to verify the effectiveness of DeepCCP compared to the state-of-the-art existing benchmark methods.

The contributions and innovations of this paper are: 1. This paper proposes an end-to-end deep learning neural network for paper citation counts prediction: DeepCCP, which avoids the process of manually extracting features and improves the prediction accuracy; 2. DeepCCP combines the information cascade effect and builds an information cascade citation network from the perspective of information cascade prediction to model paper citation prediction, providing a new idea for paper citation prediction; 3. DeepCCP only needs the structure and temporal information of the paper information cascade citation network, and does not need other information such as author characteristics and journal attributes. It uses less information and improves the prediction accuracy, which widens the universality of the scenario of paper citation prediction; 4. Our experiments on 6 data sets of 2 categories have verified the superior performance of DeepCCP compared to other baseline methods.

The rest of this article is organized as follows: In the section \ref{sec_Works}, we review the related work of literature citation prediction and information cascade prediction. In the section \ref{sec_Deepccp}, we put forward the problem, defined the problem in detail, and described the proposed DeepCCP in detail. The section \ref{experiment} is the experimental part. We describe the data set, measurement strategy and baseline method we adopted, and compare DeepCCP with the most advanced baseline. Finally, we summarized the paper in Section \ref{sec_conclusion} and elaborated on future work.

\section{Related Works}
\label{sec_Works}
\subsection{Citation Counts Prediction}
\label{sec_ccp_work}
There have been many literature studies in the existing work on how to predict the influence and success of academic papers. These documents mainly aim at different goals, such as predicting highly cited papers, predicting H-index of scholars, predicting future impact factors of journals, And predict the number of citations of academic papers.

This section mainly reviews the relevant literature that predicts the number of citations of academic papers. According to the different methods used in the paper, we can divide the existing work into the following categories: machine learning-based methods, feature-based methods, deep learning-based methods, and other methods.

\textbf{For the first category of methods based on machine learning}: This type of methods could be subdivided into the classification problem or regression problem of citation counts prediction. For classification problem, one of the main differences in this type of work is the classification criteria, such as whether the citation of the paper was higher than a certain value and divide the paper into 2 categories(\cite{fu2008models}, \cite{fu2010using}), or according to the citation counts of the paper, the paper was divided into high, medium, low three categories(\cite{ibanez2009predicting}, \cite{wang2011mining}) according to the citation counts, and even classified the paper with other indexes (e.g. h-index)(\cite{dong2015will}), and then classified the citation counts by using the relevant algorithms of machine learning, such as Support Vector Machine(\cite{fu2008models}, \cite{fu2010using}), K-Nearest Neighbor(\cite{ibanez2009predicting}), neural network and random forest algorithm(\cite{dong2015will}); For the regression problem, Support Vector Regression(SVR) was used for the five-year citation prediction task(\cite{chakraborty2014towards}). \cite{li2015trend} also used SVR to predict the total citation counts of paper in 10 years, 11 years and 12 years in two different ways. \cite{bai2019predicting} used the Gradient Boosting Decision Tree(GDBT) model in the citation counts prediction. \cite{robson2016can} used RF model to predict references in the field of environmental modeling. \cite{nie2019academic} extracted the author feature, time feature and other features, compared the K-Nearest Neighbor(KNN) algorithm, Random Forest(RF), gradient lifting decision tree(GDBT), extreme gradient lifting(XGB) and support vector machine(SVM) to verify the stability and outstanding performance of k-nearest neighbor algorithm.

\textbf{For feature-based methods}: Most of these methods classify the citation counts prediction problem as a regression problem. By extracting various features of academic papers, some regression models such as linear regression models were used to predict the citations. For example, \cite{yu2014citation} used stepwise multiple regression analysis to select appropriate features from the feature space by considering the external characteristics of the paper, the characteristics of the author, the characteristics of the publication, and the characteristics of the citation, and established a regression model to explain the relationship between citation influence and selected features; \cite{yan2011citation} used linear regression model to predict citations by extracting author ranking, social attributes, author authority, H-index and other characteristics of papers; \cite{yu2014citation} found the six most basic features of citation prediction, including citations and references of the previous two years Quantity, IF-5 (five-year impact factor), age of first citation, number of authors, and total number of citations of first authors were critical to the paper citation prediction task; \cite{lokker2008prediction} used 17 reference-related functions and three journal-related functions were used to predict the clinical papers cited in two years.

\textbf{For methods based on deep learning}: With the recent success of deep learning technology in various fields, scholars have begun to use deep learning methods to predict citation counts. \cite{ruan2020predicting} extracted six article features, two journal features, eight reference features, nine author features and five early citation features, and then fed them into a multi-layer BP neural network to predict the five-year citations of the paper; \cite{abrishami2019predicting} used Recurrent Neural Networks(RNNs) to take the counts of citation of the paper each year as input, and the output is the number of citation counts of the paper in the next K years; \cite{wen2020paper} developed the GRU-CPM model based on the GRU network (a specific form of Recurrent Neural Networks), which predicts the amount of citation counts by extracting the text features and author features of the paper, and obtained higher prediction accuracy; \cite{li2019neural} proposed a complex deep learning model to predict citation counts by combining the peer-reviewed text of the paper with other features. \cite{xu2019early} used the features of heterogeneous bibliographic networks and convolutional neural networks (CNN) to predict the number of citations of each paper in ten years, and improved the prediction accuracy by 5\% compared to the baseline models.

\textbf{For other types of methods}: Based on three basic mechanisms, namely priority attachment, aging and adaptability, \cite{wang2013quantifying} derived a dynamics model, which accurately predicted the future citation counts of a paper, and its performance is better than basic models such as logical models. Combining the hawkes process in the stochastic  process, \cite{yan2011citation} modeled the time delay effect of the papers and the trigger effects of recent citation counts to explain the popularity of the papers; \cite{manjunatha2003citation} utilized time analysis methods to construct several quarterly time series for each paper, and uses these to train a system for predicting changes in citation counts.

\subsection{Information Cascade Prediction}
\label{sec_ICP}
In this paper, we combine the information cascade effect and use the information cascade prediction to simulate the citation counts prediction, that is, utilize the information cascade diffusion to understand the increase in the citation of the paper. In the existing work on information cascade prediction, there are mainly the following categories: feature-based methods, stochastic process methods and deep learning methods.

\textbf{Feature-based methods}: This type of methods usually regards information cascade prediction problems as regression problems (\cite{szabo2010predicting}, \cite{pinto2013using}, \cite{bakshy2011everyone}, \cite{martin2016exploring}, \cite{tsur2012s}) or classification problems(\cite{shulman2016predictability}, \cite{cheng2014can}). These methods performed information cascade prediction by manually extracting multiple features. These features include temporal information(\cite{szabo2010predicting}, \cite{pinto2013using}), user characteristics\\(\cite{Cui2013Cascading}, \cite{bakshy2011everyone}, \cite{lerman2008analysis}), content characteristics(\cite{tsur2012s}, \cite{petrovic2011rt}, \cite{Ma2013On}, \cite{hong2011predicting}), and structural characteristics(\cite{romero2011interplay}, \cite{bao2013popularity}, \cite{weng2014predicting}) or mix these features(\cite{tsur2012s}), and employed these features as input to machine learning algorithms. \cite{bakshy2011everyone} revealed that user characteristics play an important role in information cascade prediction. However, some recent studies(such as \cite{shulman2016predictability}) believed that temporal features have an important influence on information cascade prediction. Nevertheless, this type of methods relied too much on features. Features are extracted by hand, which requires rich professional knowledge and has great subjectivity. Moreover, the features required for different scenes are often different, and the universality of the method is not high.

\textbf{Stochastic process method}: this kind of methods usually regards the scale of information cascade as the accumulation of messages, and models the intensity function of the process in which such messages arrive. These methods can be subdivided into the following two categories: 1) Based on Poisson process (\cite{shen2014modeling}, \cite{cho2014learning}): these methods modeled "Matthew effect" by Reinforcement Poisson process, and combined them into Bayesian framework for external factor inference and parameter estimation. 2) Based on Hawkes process (\cite{zhao2015seismic}, \cite{cao2017deephawkes}, \cite{bao2015modeling}, \cite{rizoiu2017expecting}): these methods used Hawkes self excitation point process and feature-driven method to model each cascade, and mainly estimated the content popularity, memory decay and user influence. However, these methods are usually not directly optimized for future popularity and learn the parameters of each message independently. As a result, they cannot make full use of the information implied in all cascades.

\textbf{Methods based on deep learning}: This type of methods has benefited from the recent development of deep learning, especially the significant progress of deep learning in the field of graph representation learning (\cite{perozzi2014deepwalk}, \cite{grover2016node2vec}, \cite{tang2015line}, \cite{ribeiro2017struc2vec}). This type of methods employed deep learning related algorithms to model the cascade network formed in the cascade process, then automatically and comprehensively extract the structure, temporal and other information of the cascade network in an end-to-end manner(\cite{Li2016DeepCas}, \cite{cao2017deephawkes}, \cite{chen2019information}, \cite{huang2019cascade2vec}, \cite{islam2018deepdiffuse}). DeepCas \cite{Li2016DeepCas} is the first deep learning framework for information cascade prediction. It mainly utilized the Node2Vec \cite{grover2016node2vec} method to sample sequences from the cascade network, and then puts these sequences into the neural network composed of the GRU networks and the attention mechanism; DeepHawkes \cite{cao2017deephawkes} combined the hawkes process, and then modeled the cascade network through the GRU network to enhance the interpretability of the information cascade prediction; CasCN \cite{chen2019information} sampled the subgraphs in the cascaded network, and then fed the subgraphs into the convolutional neural networks and GRU networks, which greatly improves the prediction accuracy. Compared with other types of methods, this type of deep learning methods could extract and process a variety of information well, and then achieve a good prediction effect.

In summary, in order to overcome the shortcomings of the various methods of citation prediction, we are inspired by the information cascade prediction methods based on deep learning, and rely on the existing deep learning paradigm to develop an end-to-end deep learning framework: DeepCCP. The main idea is to analogize the paper citation process to the information cascade process, and treat the paper citation network as an information cascade network, utilize the deep learning method to model the citation network, and extract the structure and temporal information of the citation network. And after a large number of comprehensive experiments, the experimental results verify the effectiveness of DeepCCP.

\section{DeepCCP: Model Overview}
\label{sec_Deepccp}
\subsection{Problem Definition}
\label{sec_Define}
Citation prediction is a very complicated issue. In order to have a deeper discussion and facilitate understanding, in this section, we elaborate on this issue and define some key concepts of DeepCCP.

\begin{figure}
\centering
\includegraphics[width=\textwidth]{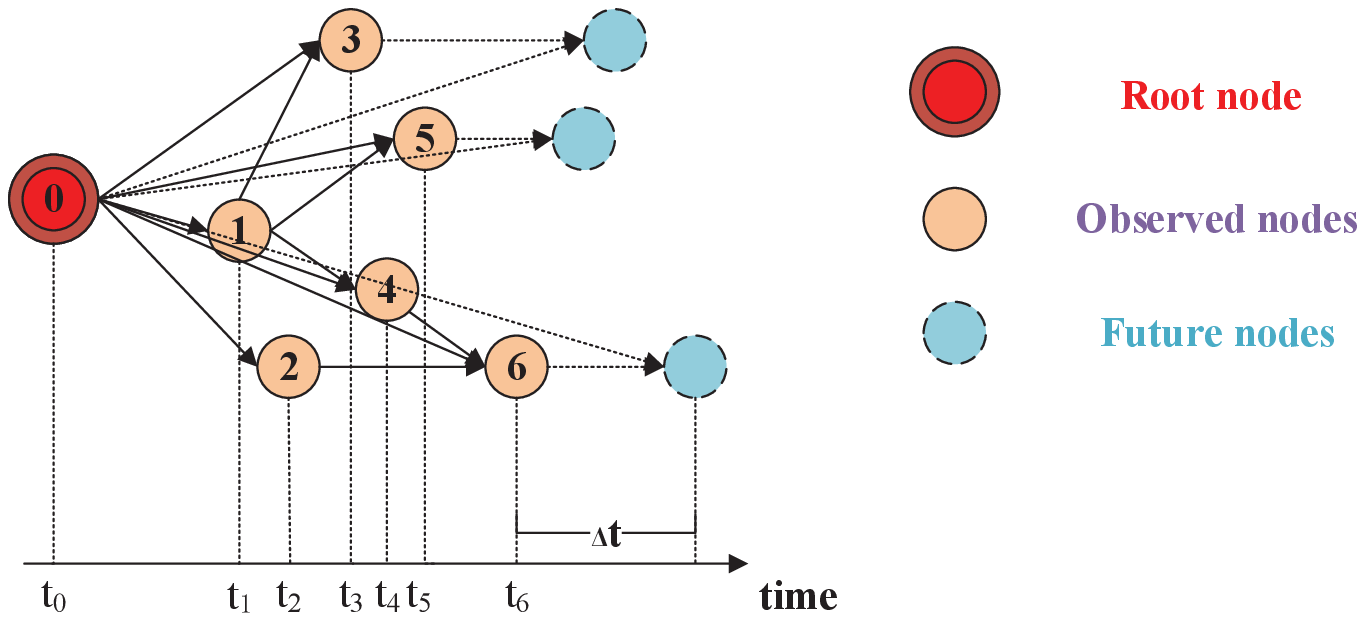}
\caption{An Example of Citation Network.} \label{cascade}
\end{figure}

$\mathbf{Definition}$ 1: \emph{Citation Network/Cascade Network}. A citation network is a network generated by mutual citations between papers. We think it is also a kind of cascade network. Given a set of papers $P$, we use $p_{i}$ to represent the i-th paper in the set $P$. For the paper $p_{i}$, the citation network $C_{i}^{T}$ formed during the observation time $T$ can be defined as $C_{i}^{T}$=$ [(v_{1},t_{1}, e_{i}(t_{i})), (v_{2},t_{2}, e_{i}(t_{i}))$,\\$ \ldots (v_{m},t_{m}, e_{i}(t_{i}))]$, Where $v_{m}$,$t_{m}$ represents the node $v_{m}$ participating in this citation network at time $t_{m}$, and $e_{i}(t_{i})$ represents the set of citation relationships between paper $v_{m}$ and paper $v_{i}$ at time $t_{i}$. According to this definition, we can also conclude that citation networks is a directed acyclic graphs(DAGs), as shown in Figure~\ref{cascade}.

According to the citation network in the Figure~\ref{cascade}, we can get different citation network representations according to different observation times t, such as:$C^{t_0}$=$[(0,t_0,(\emptyset))]$, $C^{t_1}$=$[(0,t_0,((0,1)))$,$(1,t_1,(\emptyset))]$, $C^{t_2}$=$[(0,t_0,((0,1)$,\\$(0,2)))$,$(1,t_1,(\emptyset))$,$(2,t_2,(\emptyset))]$, $\ldots$,
$C^{t_6}$=$[(0,t_0,((0,1),(0,2),(0,3),(0,4),(0,5)$,\\$(0,6)))$,$(1,t_1,((1,3)$,$(1,4)$,$(1,5)))$,$(2,t_2,((2,6)))$,$(3,t_3,(\emptyset))$,$(4,t_4,((4,6)))$,$(5,t_5$,\\$(\emptyset))$,$(6,t_6,(\emptyset))]$

Such an explanation seems to be no problem, but a closer look at the figure \ref{cascade} shows that a citation network is not a standard tree structure. A paper will cite many papers in the citation network, that is, a node may have more than one parent node. However, the input of DeepCCP requires a tree structure, so a certain transformation of the citation network is required. The transformation process will be explained in detail in the section \ref{experiment}.

$\mathbf{Definition}$ 2: \emph{the Growth of Citation Counts}. Given a citation network $C$ and the observation time $t$, its subgraph at time $t$ is denoted as $C^t$, and the number of nodes is denoted as $\left| V_ {t} \right|$. After the time $\delta$$t$, the number of nodes in the cascade network $C_{t+\Delta t}$ is denoted as $\left| V_{t+\Delta t} \right|$, then the Growth of Citation Counts from $t$ to $t+\Delta t$ is $\left| V_{t+\Delta t} \right|$-$\left| V_{t} \right|$, denoted as $G_{C}^{\Delta t}$.

$\mathbf{Definition}$ 3: \emph{Network Hierarchy}. Given a citation network $C$, the $k$-th hierarchy of this citation network refers to the set of nodes with the shortest distance $k$ from the root node of the citation network, denoted as $V_{C}^{k}$, where $V_{C}^{0}$ refers to the root node of the cascade network. For example, in the figure \ref{cascade}, the first hierarchy $V_C^1$=$\{1,2,3\}$.

$\mathbf{Definition}$ 4: \emph{The Maximum Depth of the Network}. The maximum depth of a citation network can be understood by using the concept of \emph{Network Hierarchy}. For a citation network $C$, find all its network hierarchies $V_{C}^{k}$ in order from $k = 1$. When a certain value of $k$ satisfies $V_{C}^{k} \ne \emptyset$  and $V_{C}^{k+1} = \emptyset$ , the maximum depth of the citation network is \emph{k}.

\subsection{Model Design}
\label{sec_Design}
\begin{figure}
\centering
\includegraphics[width=\textwidth]{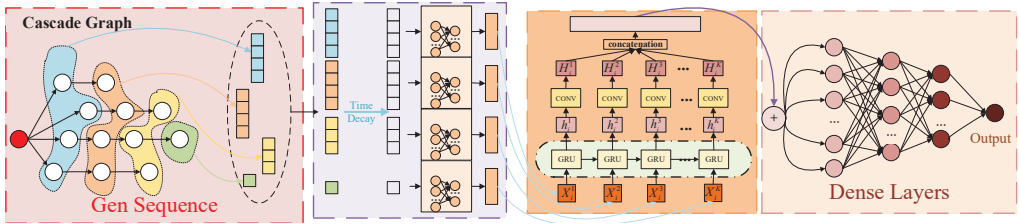}
\caption{The Framework of DeepCCP.} \label{framework}
\end{figure}

The deep learning framework DeepCCP proposed in this paper takes a citation network as input and predicts the citation growth of the root node (paper) of this citation network in a period of time $t$. DeepCCP mainly uses the degree sequence to encode the primary structure of the citation network, and then utilizes recurrent neural network (GRU) and convolutional network to further capture the potential structure and temporal information of the citation network. The framework of DeepCCP is shown in the figure \ref{framework}. Below, we will briefly describe the general structure of this framework, and describe each part of the framework in detail in the following sections.

As an end-to-end deep learning framework, DeepCCP is mainly composed of the following 3 parts.
(1) Generating the degree sequences of citation networks. This is the first step of DeepCCP, which mainly divides the degree sequence of the citation network according to the hierarchy of the citation network. The degree sequence contains structural information of the citation network, which can preliminarily represent the citation network.
(2) Extracting the structure and time information of the citation network. This is a crucial step in DeepCCP, because it can dig out the complex structural information and temporal features of the citation network, which can help us achieve the best results in the application of citation counts predicting. This part is mainly composed of four modules: a structural hidden layer composed of MLP(Multilayer Perceptron), time coding layer, GRU network and convolutional network. These modules will be explained in detail in subsequent sections.
(3) Predict the growth of paper citation counts through Multilayer Perceptron (MLP). The low-dimensional representation of the citation network obtained in (2) can be used as the input of MLP to train the model to predict the growth of paper citation counts.

\subsubsection{Generating the Degree Sequence of the Citation Network}
\label{sec_generate}
The first step of DeepCCP is to generate the degree sequences of the citation networks.
As we all know, as a nonlinear topology, the network is a complex object, and traditional machine learning methods cannot directly process it. Therefore, in order to facilitate processing, the network is usually converted into an adjacency matrix and other similar forms that can be recognized by computers. However, the adjacency matrix cannot reflect the hierarchical structure information of the network well. In order to solve this defect, we proposed a novel representation method of the citation networks: degree sequence. We will elaborate on this method below.
\begin{figure}
\centering
\includegraphics[width=\textwidth]{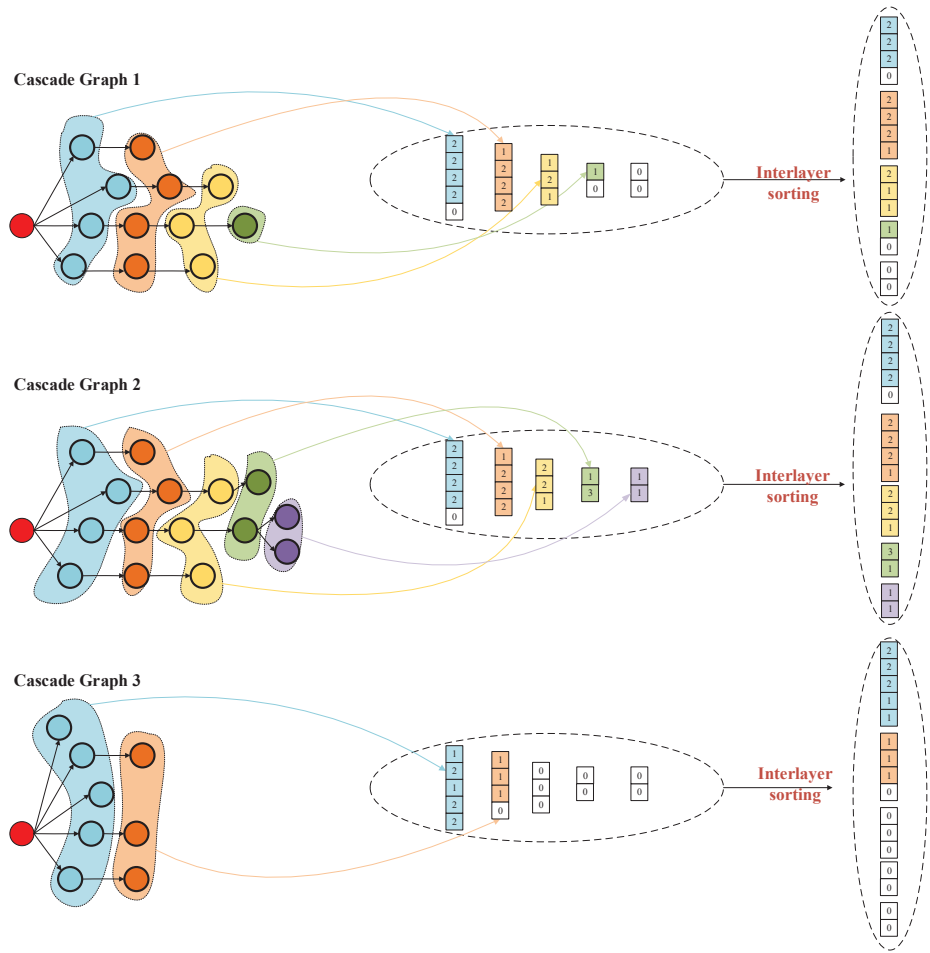}
\caption{Generating the Degree Sequence of the Citation Network. } \label{primary}
\end{figure}
Given $n$ citation networks, to generate the degree suquence of each citation network, the first thing to do is to determine the length of the degree suquence of the citation network $C_{i}$, It needs to be added that the degree sequence is divided according to the network level, and it is composed of multiple subsequences, so we need to determine the length of each subsequence. it can be described in mathematical language as:
\begin{equation}
L_{X}^k = \max (\left| V_{C_{1} }^{k} \right| ,\left| V_{C_{2} }^{k} \right| ,\ldots \left| V_{C n}^{k} \right| ),k\in[1,K]
\label{LX}
\end{equation}
$L_{X}^k$ represents the length of the $k$-th subsequence of degree sequence, $K$ represents the maximum value of the maximum depth of $n$ citation networks, max($\cdot$) represents the maximum value in the sequence $\cdot$, $\left| V_{C_{i} }^{k} \right|$ represents the number of nodes at the $k$-th hierarchy  in the citation network..

Second, we have to determine the value of the degree sequence. In the previous step we have clarified the length of each subsequence in the degree sequence, the length of each subsequence is denoted as $L^{1}$,$L^{2}$,$L^{3}$,...,$L^K$. For the citation network $C_{i}$, the value of different subsequence is the degree of corresponding hierarchy nodes in this citation network. If the number of nodes in the hierarchy is less than the length of the corresponding subsequence, then we pad it with 0. We believe that the greater the locality, the greater influence the node has on the future citation. In order to highlight this effect, we do hierarchical ordering and arrange the points with large locality in the front, when there are multiple nodes with the same degree, we sort the nodes twice based on the temporal information. So far, the degree sequence $X_{i}$ of the citation network $C_{i}$ is completed, where $X_{i}^k$ represents the $k$-th subsequence.

\begin{figure}
\centering
\includegraphics[width=\textwidth]{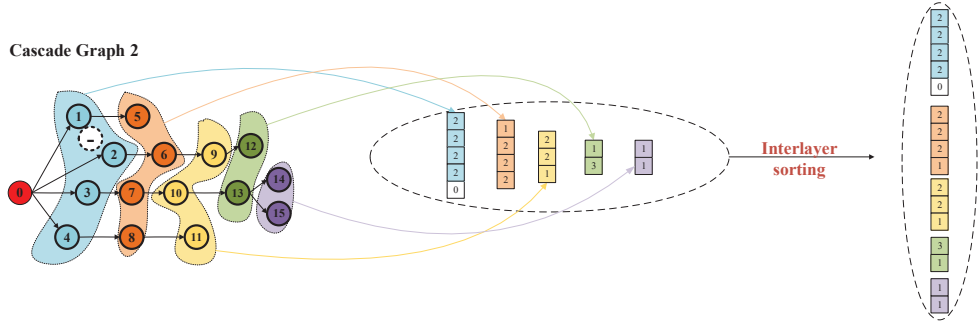}
\caption{Citation Network Hierarchy and Its Degree Sequence} \label{primary_c2}
\end{figure}

Let's take Figure~\ref{primary} as an example to show how to construct the degree sequence of a citation network. First of all, we can get the three citation networks, citation network 1 (hereinafter to as $G1$), citation network 2 (hereinafter to as $G2$) and citation network 3 (hereinafter to as $G3$). The depth of the 3 corresponding network are respectively
4, 5 and 2. Obviously, We can define $K$ = 5, which means that every citation network degree sequence contains 5 subsequences, and $L_1$=$\max (\left| V_{G_{1} }^{1} \right| ,\left| V_{G_{2} }^{1} \right| ,\left| V_{G_3}^{1} \right| ) = 5$, similarly $L_2 = 4$, $L_3 = 3$, $L_4 = 2$, $L_5 = 2$, then $L_X = \sum_{k=1}^{5}{L_k}=16$. We will show the process of constructing the degree sequence for $G2$ in Figure~\ref{primary_c2}.

It is noted that there are only 15 nodes in addition to the root node in $G2$, But the total length of each subsequence is 16. That is because that the maximum size of $|V_{G_i}^1|$ is 5 and $G2$ only has 4 nodes in the first layer. Thus, we can add an inactive virtual node in the first level of $G2$, as shown in Figure~\ref{primary_c2}, and it can also be applied to $G1$ and $G3$.

The process of obtaining the degree sequence is shown in Algorithm ~\ref{alg1};

\begin{algorithm}
\caption{Generating Degree Sequence of Citation Networks}
\label{alg1}
\begin{algorithmic}[1]
\REQUIRE ~~ $Citation$ $Network$ $set$ $C=(C_1,C_2,\ldots,C_n).$
\ENSURE $the$ $degree$ $sequences$ $X=(X_1,X_2,\ldots,X_n)$ $of$ $the$ $citation$ $network$ $set.$
\STATE $For$ $the$ $citation$ $network$ $set$ $C,$ $find$ $the$ $maximum$ $depth$ $of$ $each$ $network$ $and$ $find$ $the$ $maximum$ $value$ $K$ $among$ $them;$
\FOR{$each\,citation\,network\,C_i$}
\STATE $Find$ $the$ $1-K$ $level$ $nodes$ $V_{C_i}^k$ $and$ $the$ $number$ $of$ $nodes$ $\vert V_{C_i}^k \vert$ $of$ $the$ $citation$ $network$ $C_i$;
\ENDFOR
\STATE $Find$ $the$ $maximum$ $number$ $of$ $nodes$ $(L_1,L_2,\ldots,L_K)$ $in$ $the$ $1-K$ $hierarchy$ $of$ $cascade$ $networks;$
\STATE $Calculate$ $the$ $subsequence$ $length$ $of$ $degree$ $sequence$ $by$ $Equation$ (~\ref{LX});
\FOR {$each\,citation\,network\,C_i$}
\STATE $Construct$ $K$ $sequences. $ $The$ $corresponding$ $positions$ $are$ $filled$ $into$ $the$ $degree$ $of$ $the$ $corresponding$ $nodes$ $according$ $to$ $the$ $points$ $of$ $the$ $K$ $hierarchies$ $of$ $the$ $citation$ $network$ $C_i, $ $and$ $insufficient$ $0s$ $are$ $filled;$
\STATE $Sort$ $between$ $hierarchies;$
\ENDFOR
\end{algorithmic}
\end{algorithm}

From Figure~\ref{primary_c2} we can know that $G2$ has 5 levels: $V_{G2}^{1}=(1,2,3,4)$,$V_{G2}^{2}=(5,6,7,8)$, $V_{G2}^{3}=(9,10,1 1)$, $V_{G2}^{4}=(12,13)$, $V_{G2}^{5}=(14,15)$. Then we need to construct 5 sequences. According to the processes above, the 5 sequences are (1,2,1,2,-), (1,2,2,2), (2,2,1), (1,3), (1,1). Since the first sequence has the limitation of $\left| V_{G2}^{1} \right|$=4$\neq$$L_1$, it is necessary to add (L1-$\left| V_{G2}^{1} \right|$=1) 0, the first sequence becoming (1,2,1,2,0). Then, in order to highlight the influence of the nodes, we sorted these 5 sequences separately, the original 5 subsequences become: [(2,2,2,2,0), (2,2,2,1), (2,2,1), (3,1), (1,1)], which is the degree sequence of $G2$.

\subsubsection{Extracting Potential Structural and Temporal Information}
\label{sec_extract}
For any citation network, all nodes in the network have different time characteristics, such as when the root node of the citation network is published and when other papers cite other papers in the citation network. We can use the cascade effect to understand this phenomenon: Generally, the earlier a node joins the cascade, the greater the impact on the entire cascade may be. On the contrary, the later the node appears in the cascade, the smaller the impact on the cascade. This is what we call the time decay effect. Paper citation is also a classic cascade phenomenon, and the time information of nodes is equally important. Therefore, the framework DeepCCP proposed in this paper requires critical time information to learn this time decay effect. However, most of the existing researches directly define some time decay functions based on professional knowledge~\cite{mishra2016feature}, such as power law time decay functions or exponential time decay functions. However, for different situations, especially some unknown situations, it is impossible to determine which type of time decay function to use, which is full of uncertainty.
\begin{figure}
\centering
\includegraphics[width=\textwidth]{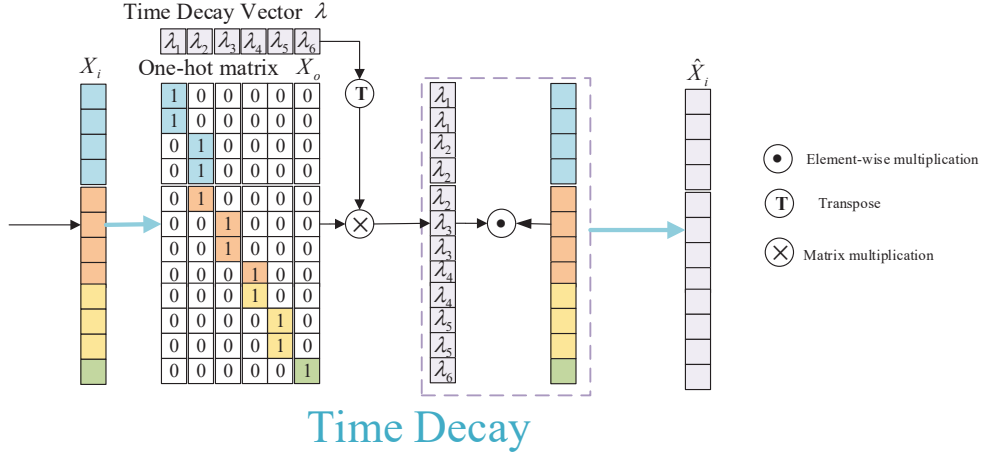}
\caption{Time Decay Effect} \label{time_decay}
\end{figure}

In this paper, we decided to adopt the method mentioned in~\cite{cao2017deephawkes} to directly learn the time decay effect as shown in Fig.~\ref{time_decay}. The specific approach is as follows. Suppose we start observation from time 0 until the end of time $T$. During time $T$, the time decay effect has been continuously changing. Now we divide the length of time range $T$ into $L$ disjoint intervals: $\{[t_0=0,t_1),[t_1,t_2),\ldots,[t_{L-1},\\t_L=T)\}$ and learn the corresponding discrete variables of the time decay effect: $\lambda _{l}$, $l \in$ (1, 2,...,L) to approximate this time decay effect function. The mapping function \emph{f} from continuous time to time interval is defined as:
\begin{equation}
\label{project_time}
f(t_{i}^{j} )=l,~ if \quad t_{l-1} \le t_{i}^{j} < t_{l}
\end{equation}
$t_{i}^{j}$ represents the timestamp when the $j$-th node of the cascade network $C_{i}$ is included in the cascade.

In the last section, we have obtained the degree sequence $X_{i}$ of the citation network $C_{i}$, ten the citation network $C_{i}$ and time characteristics are combined in the following way: First, convert each element in $X_{i}$ into a one-hot vector through Equation \ref{project_time}. The index of the position of 1 is the value of the $f$ function. Then the one-hot matrix $X_{o}$ is multiplied by time decay effect parameters $\lambda$ to get $X_{\lambda}$, finally $X_{\lambda}$ and $X_{i}$ are dot-multiplied. The whole process is shown in the Fig.~\ref{time_decay} and can be expressed in mathematical language as:

\begin{equation}
\label{TIMEx}
\begin{split}
X_{o} &= Onehot(X_{i})\\
X_{\lambda} &= \lambda \times X_{o} \\
\tilde{X_{i}} &=X_{\lambda} \cdot X_{i}
\end{split}
\end{equation}
Where $Onehot(\cdot)$ is a function that transforms $\cdot$ into a one-hot vector, and $\tilde{X_{i}}$ represents the degree sequence integrated with temporal information.

After integrating the temporal information, the next step is to initially extract the time and structure information contained in $\tilde{X_{i}}$. Our approach is to introduce a pre-embedding layer, which is a neural network composed of multiple fully connected layers. Another function of the pre-embedding layer is to convert each subsequence of the degree sequence into a sequence of equal length, so that the degree sequence can be used as the input of the subsequent neural network. Just feed the sequences $\tilde{X_{i}}$ obtained by the Equation \ref{TIMEx} to this pre-embedding layer. It can be expressed by the following equation:

\begin{equation}
\hat{X_i^k} = PreEmbedding(\tilde{X_{i}^k}),k\in[1,K]
\label{preembedding}
\end{equation}

In order to further obtain more dense and high-quality structural information of citation networks, We use GRU network and 1D convolutional neural network.
Firstly, we use Gated Recurrent Unite(GRU)\cite{1997Long}, a specific type of recurrent neural network(RNN), which is known to be effective for modeling sequences. When GRU is applied to a sequence recursively from left to right, the sequence representation will be increasingly enriched by the information of subsequent sequences in the degree sequence, and the gating mechanism determines the amount of new information to be added and the amount of history to be retained , it simulates the process of information flow in the diffusion process. And our degree sequence could characterize the hierarchical structure of the citation networks. We put the sequences from the first hierarchy to the $K$-th hierarchy into the GRU in turn, and we can get a sequence rich in citation network structure information, and, because we have been in temporal information is integrated in the degree sequence, so the final sequence we get is rich in structure and temporal information. Specifically, step $k$ is represented as the $k$-th subsequence in the degree sequence in turn. For each step $k$ that takes the input subsequence $\hat{X_i^k}\in R^M$ and the previous hidden state $h_i^{k-1}\in R^M$ as input, the GRU calculates the updated hidden state $h_i^{k}$ = $GRU(\hat{X_i^k}, h_i^{k-1})$,$h_i^{k}\in R^M$:
\begin{equation}
\label{GRU}
\begin{split}
  u_i^k &= \sigma(W^{u}\hat{X_i^k}+U^{u}h_i^{k-1}+b^{u}), \\
  r_i^k &= \sigma(W^{r}\hat{X_i^k}+U^{r}h_i^{k-1}+b^{r}), \\
  \hat{h_i^k} &= tanh(W^{h}\hat{X_i^k}+U^{h}h_i^{k-1}+b^{h}),\\
  h_i^k &= u_i^k \cdot \hat{h_i^k}+(1-u_i^k) \cdot h_i^{k-1},
\end{split}
\end{equation}
where $\sigma{\cdot}$ is the sigmoid activation function, $\cdot$ represents an element-wise product. $W^{u}$, $W^{r}$, $W^{h}$, $U^{u}$, $U^{r}$, $U^{h}$ $\in R^M\times M$ and $b^{u}$, $b^{r}$, $b^{h}$ $\in R^M$ are GRU parameters, $M$ is the size of the last fully connected layer of the pre-embedded layer..

After using the GRU network to capture the deep structure information and temporal information of the citation networks, we then utilize the 1D convolutional neural network to perform convolution operations on each subsequence of output sequence $h_i$ of GRU, aiming to further extract the information inside the subsequence and capture the internal dependency. The specific operations are as follows:

\begin{equation}
H_i^k = f(W\otimes h_i^k+b)
\label{CONV1D}
\end{equation}
where $\otimes$ denotes the 1D convolution operation, $W$, $b$ are the parameter of the convolutional networks, $H_i^k \in R^{(M/2)}$.

So far, we have obtained the sequence set \{$H_i^1$, $H_i^2$, $H_i^3$, $\ldots$, $H_i^K$\}, which is deeply processed by the degree sequence. \{$H_i^1$, $H_i^2$, $H_i^3$, $\ldots$, $H_i^k$\} is rich in the structure and time information of the citation network. Before putting it in the last prediction module, what we need to do in advance is to put them connection:

\begin{equation}
H_i = Concat(H_i^1, H_i^2,\ldots H_i^K)
\label{Concatenation}
\end{equation}
Where $Concat(\cdot)$ means concatenation, $H_i \in R^{K\times M/2}$.

\subsubsection{Predicting the Growth Scale}
\label{sec_predict}
The last module of the DeepCCP is the prediction module. The output of the 1D convolutional network after the Equation~\ref{Concatenation} is used as the input of the MLP, and the output of MLP is the predicted growth of citation counts ${\mathop {G_{i}^{\Delta t} }\limits ^{ \tilde{\ }}}$ of the citation network , the process is described as follows:
\begin{equation}\label{MLP}
{\mathop {G_{i}^{\Delta t} }\limits ^{ \tilde{\ }}} =MLP(H_i)
\end{equation}
Our final task is to predict the growth of citation counts of the fixed time interval \emph{t} of the citation network, which can be accomplished by minimizing the following loss function:
\begin{equation}\label{loss_mlp}
L_{pre} =\frac{1}{N} \sum\limits _{i=1}^{N}(\log _{2} G_{C_{i} }^{\Delta t} -\log _{2} {\mathop {G_{C_{i} }^{\Delta t} }\limits ^{ \tilde{\ }}} )^{2}
\end{equation}
$N$ represents the number of citation networks in training set with ground-truth labels of growth citation count.

So far, the DeepCCP models we have proposed jointly minimize the following objective loss functions:
\begin{equation}\label{loss-deepccp}
  \begin{split}
     L_{DeepCCP} &=\alpha L_{pre} +\beta L_{reg} \\
       &==\frac{\alpha }{N} \sum\limits _{i=1}^{N}(\log _{2} G_{C_{i} }^{\Delta t} -\log _{2} {\mathop {G_{C_{i} }^{\Delta t} }\limits ^{ \tilde{\ }}} )^{2} +\beta L_{reg}
  \end{split}
\end{equation}
$\alpha$, $\beta$ are the tradeoff parameters. $L_{reg}$ is a L2 norm regularization term to avoid over-fitting, defined as follows:
\begin{equation}\label{l2w}
  L_{reg} =\sum\limits _{i=1}^{M}\left\| W^{(i)} \right\| _{F}^{2}
\end{equation}

From the optimization objective in Equation~(\ref{loss-deepccp}), we can see that besides that the training citation networks with ground-truth labels is used to minimize the prediction error in $L_{DeepCCP}$. The algorithm description of the whole process is shown in Algorithm~\ref{alg2}.
\begin{algorithm}
\caption{DeepCCP}
\label{alg2}
\begin{algorithmic}[1]
\REQUIRE ~~ $The$ $degree$ $sequence$ $set$ $X=(X_1,X_2,\ldots,X_n)$ $of$ $the$ $citation$ $networks,$ $the$ $time$ $information$ $T$ $of$ $the$ $citation$ $networks.$
\ENSURE $The$ $predicted$ $scitation$ $counts$ $of$ $the$ $citation$ $network$ $set$ $GA=(\mathop {G_{C_{1} }^{\Delta t} },\mathop {G_{C_{2} }^{\Delta t} },\ldots,\mathop {G_{C_{n} }^{\Delta t} }).$
\REPEAT
\FOR{$each\,citation\,network\,C_i$}
\STATE $The$ $degree$ $sequence$ $X_i$ $of$ $the$ $citation$ $network$ $C_i$ $obtains$ $the$ $vector$ $\hat{X_i}$ $of$ $the$ $integration$ $time$ $information$ $according$ $to$ $Equations$ $(\ref{project_time}),$ $(\ref{TIMEx})$ $and$ $(\ref{preembedding})$;
\STATE $Vector$ $\hat{X_i}$ $is$ $fed$ $to$ $the$ $GRU$ $network$ $as$ $input$, $getting$ $vectors$ ${h_i^k}$ $that$ $rich$ $in$ $temporal$ $and$ $structure$ $information$ $through$ $equation$ $\ref{GRU};$
\STATE $feeding$ $vectors$ ${h_i^k}$ $into$ $1D$ $convlutional$ $networks,$ $then$ $calculate$ $H_i$  $by$ $Equations$ $\ref{CONV1D}$ $and$ $\ref{Concatenation};$
\STATE $Calculating$ $the$ $predicted$ $growth$ $citation$ $counts$ $\mathop {G_{C_{i} }^{\Delta t} }$ $of$ $the$ $citation$ $network$ $C_i$ $by$ $Equation$~(\ref{MLP});
\STATE $Using$ $Adaptive$ $moment$ $estimation$ $(Adam)$ $to$ $optimize$ $the$ $objective$ $function$ $in$ $Equation$ $~\ref{loss-deepccp}, $ $and$ $update$ $the$ $parameters$ $in$ $Equations$ (\ref{TIMEx}),(\ref{preembedding}),~(\ref{GRU}),~(\ref{CONV1D}),~(\ref{MLP});
\ENDFOR
\UNTIL{convergence;}
\end{algorithmic}
\end{algorithm}

\section{Experiment}
\label{experiment}
In this section, we compare the performance of our proposed model DeepCCP with several state-of-the-art approaches that we use as baselines, for citation count prediction using two real-world datasets APS dataset and HEP-PH dataset.

\subsection{Datasets}
\label{sec_Datasets}
\begin{figure}
  \centering
  \includegraphics[width=\textwidth]{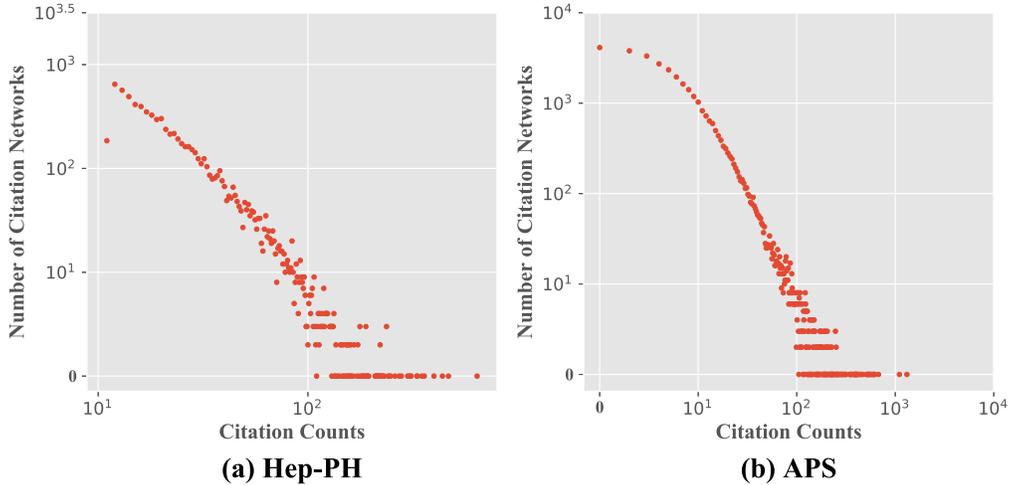}
  \caption{Distribution of Citation Counts. }\label{NUMBERS POPU}
\end{figure}

We evaluate the effectiveness and generality of DeepCCP in two public citation datasets and compare it with previous work such as DeepCas and DeepHawkes.The first one is to predict the citation counts of all papers and citations from 11 journals of American Physical Society between 1893 and 2009, the second one is to predict the citation count of papers in e-print arXiv and covers papers in the period from January 1993 to April 2003 (124 months). The statistics of the datasets as shown in Table ~\ref{statistics}. Figure ~\ref{NUMBERS POPU} shows the relationship between the number of cascade graphs and the growth scale of the two datasets.

\textbf{HEP-PH\cite{chen2019information}}: The first data set is HEP-PH, which is from E-print arXiv and covers the papers from January 1993 to April 2003 (124 months). If paper I refers to paper J, the graph contains directed edges from I to J. This data was originally released as part of the 2003 KDD cup [57]. For the observation window, we chose t = 3, 4, and 5 years, corresponding to the years when the epidemic reached 50\%, 60\% and 70\% of the final size, as shown in Figure 5 (b). Then, we select 70\% of the citation graphs for training, and the rest are verified and tested by bisection.

\textbf{APS\cite{shen2014modeling}}: We now turn to the scenario of predicting the citation count of papers. The dataset used in this paper is from American Physical Society (APS) [9], including all the papers published by the 11 APS journals between 1893 and 2009, and the citations among these papers. In this scenario, the unit of time is day, and for each citation we record the number of days elapsed since the publication of the cited paper. All the citations to a paper form a cascade, and the popularity of cascade is the number of citations.

\begin{figure}
  \centering
  \includegraphics[width=\textwidth]{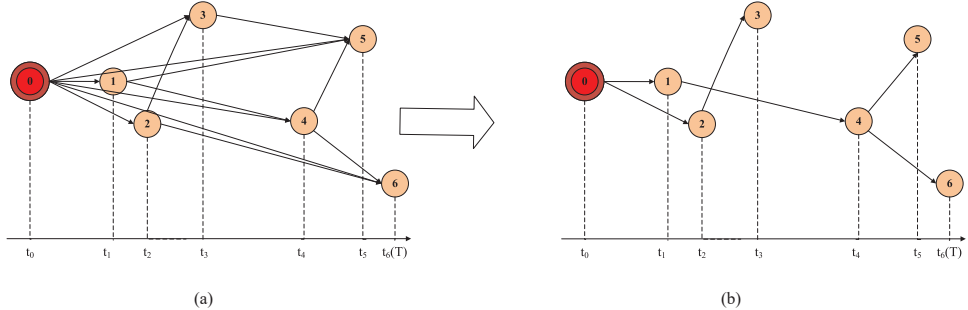}
  \caption{ Illustration of Citation Network Construction. }\label{trans}
\end{figure}

Earlier we mentioned in Definition 1 that for the real scenario of a citation network, it may not be a standard tree structure. For example, in a citation network, a paper has more than 2 references, or it is forwarded In the network, a user can forward the posts of more than two users at the same time, which leads to a network of more than two parent nodes on a node, that is, the hierarchical structure is not a tree structure, which is not conducive to our operation. To this end, we follow the method mentioned in ~\cite{cao2017deephawkes} to solve the implicit cascading diffusion path of each node. The specific solution is to keep only the connection of the parent node with the latest participation time among all parent nodes when there are more than two parent nodes in a node, and delete its connection with other parent nodes. Take the citation network in figure ~\ref{trans}(a) as an example, we can see that node 6 has two diffusion paths: "0-1-4-6" and "0-2-6", namely nodes 6 has 4 and 2 parent nodes, but as shown in the figure ~\ref{trans}(a), 4 joined the citation network later than 2, so only the connection between 6 and 4 is kept and 6 and The connection between 2, that is, the diffusion path of node 6 is only: "0-1-4-6", the citation network converted in figure ~\ref{trans}(a) is shown in figure~\ref{trans}(b).

\begin{table}[]
\caption{Statistics of Datasets}\label{statistics}
\centering
\begin{tabular}{c|c|c|c|c|ccc}
\hline
                                 & Dataset & \multicolumn{3}{c|}{Hep-PH}  & \multicolumn{3}{c}{APS}                                            \\ \hline
papers                           & All     & \multicolumn{3}{c|}{34,546}  & \multicolumn{3}{c}{463, 348}                                       \\ \hline
edges                            & All     & \multicolumn{3}{c|}{421,578} & \multicolumn{3}{c}{4, 710, 547}                                    \\ \hline
T                                &         & 3y       & 4y      & 5y      & \multicolumn{1}{c|}{5y}     & \multicolumn{1}{c|}{7y}     & 9y     \\ \hline
\multirow{3}{*}{Numbers}         & train   & 4,978    & 5,481   & 5,748   & \multicolumn{1}{c|}{16,299} & \multicolumn{1}{c|}{21,171} & 24,658 \\
                                 & val     & 1,067    & 1,175   & 1,232   & \multicolumn{1}{c|}{3,582}  & \multicolumn{1}{c|}{4,507}  & 5,254  \\
                                 & test    & 1,066    & 1,174   & 1,231   & \multicolumn{1}{c|}{3,475}  & \multicolumn{1}{c|}{4,589}  & 5,279  \\ \hline
\multirow{3}{*}{Avg.Path Length} & train   & 3.033    & 3.052   & 3.072   & \multicolumn{1}{c|}{2.12}   & \multicolumn{1}{c|}{2.28}   & 2.41   \\
                                 & val     & 3.069    & 3.088   & 3.101   & \multicolumn{1}{c|}{2.07}   & \multicolumn{1}{c|}{2.27}   & 2.41   \\
                                 & test    & 3.158    & 3.301   & 3.428   & \multicolumn{1}{c|}{2.14}   & \multicolumn{1}{c|}{2.3}    & 2.42   \\ \hline
\multirow{3}{*}{Avg.Popularity}  & train   & 3.7      & 3.9     & 4       & \multicolumn{1}{c|}{17.8}   & \multicolumn{1}{c|}{19.2}   & 20.2   \\
                                 & val     & 3.2      & 3.4     & 3.7     & \multicolumn{1}{c|}{17.3}   & \multicolumn{1}{c|}{18.9}   & 20     \\
                                 & test    & 3.5      & 3.6     & 3.8     & \multicolumn{1}{c|}{17.9}   & \multicolumn{1}{c|}{19.1}   & 20.3   \\ \hline
\multirow{3}{*}{Avg.Degree}      & train   & 2.000    & 2.000   & 2.000   & \multicolumn{1}{c|}{2.000}  & \multicolumn{1}{c|}{2.000}  & 2.000  \\
                                 & val     & 2.000    & 2.000   & 2.000   & \multicolumn{1}{c|}{2.000}  & \multicolumn{1}{c|}{2.000}  & 2.000  \\
                                 & test    & 2.000    & 2.000   & 2.000   & \multicolumn{1}{c|}{2.000}  & \multicolumn{1}{c|}{2.000}  & 2.000  \\ \hline
\multirow{3}{*}{Avg.Edge}        & train   & 4.270    & 4.270   & 4.270   & \multicolumn{1}{c|}{18.875} & \multicolumn{1}{c|}{20.773} & 22.949 \\
                                 & val     & 3.310    & 3.930   & 3.950   & \multicolumn{1}{c|}{18.667} & \multicolumn{1}{c|}{20.542} & 22.873 \\
                                 & test    & 3.910    & 3.270   & 3.280   & \multicolumn{1}{c|}{19.844} & \multicolumn{1}{c|}{22.236} & 24.707 \\ \hline
\multirow{3}{*}{Avg.Leaf node}   & train   & 2.161    & 2.400   & 3.007   & \multicolumn{1}{c|}{11.304} & \multicolumn{1}{c|}{12.196} & 13.331 \\
                                 & val     & 2.283    & 2.636   & 3.265   & \multicolumn{1}{c|}{11.071} & \multicolumn{1}{c|}{11.931} & 13.140 \\
                                 & test    & 2.351    & 2.904   & 3.404   & \multicolumn{1}{c|}{11.717} & \multicolumn{1}{c|}{12.879} & 14.157 \\ \hline
\end{tabular}
\end{table}

\subsection{Baselines}
\label{sec_base}
\textbf{Feature-Based}: Recent studies have shown that structural features are informative for information cascade prediction. In our paper, we extract some of the above-mentioned features from the cascade graph, and use them to predict the citation counts through two ways: \textbf{Feature-Linear} and \textbf{Feature-Deep}, the label of each citation networks has adopted the way of logarithmic transformation. For \textbf{Feature-Linear}, We feed the features into a linear regression model; and for \textbf{Feature-Deep}, we feed the hand-craft features into a MLP model to predict the citation counts of citation networks.

\noindent\textbf{Node2Vec}\cite{grover2016node2vec}: Node2Vec is selected as a representative of node embedding methods,it is an extension of DeepWalk. We perform random walk for each cascade graph and get the embedding vector of each node, then we feed the embedding vectors of nodes that from the same cascade graph into a MLP model to get citation counts of different citation networks. In this way, Node2Vec can extract the potentially complex structure of the citation network, so it can better predict the citation counts of the paper.

\noindent\textbf{DeepCas}\cite{Li2016DeepCas}: DeepCas is the first deep learning architecture for information cascade prediction of two types of cascades in social and paper citations, it samples a set of nodes paths by random walk and feeds paths into a bi-directional GRU neural network with an attention mechanism to predict the growth size of the cascade graph.DeepCas mainly utilizes the structure of the cascade graph and node identities.

\noindent\textbf{DeepHawkes}\cite{cao2017deephawkes}: DeepHawkes integrates the interpretability of hawkes process and the high performance of information cascade prediction. Therefore, it also has a good performance in the citation counts prediction scenario. Deepawkes integrates the interpretability of hawkes process and the high performance of information cascade predictor. It models the dynamic formation process of cascade graph by hawkes process, and establishes the bridge between prediction and information cascade prediction.

\noindent\textbf{Topo-LSTM}\cite{wang2017topological}: Topo-LSTM is a novel recurrent neural network with topological structure. It takes a directed acyclic graph as input and generates an embedding vector for each node in the graph.

\noindent\textbf{CasCN}\cite{chen2019information}: CasCN is a novel multi-directional/dynamic GCN, it samples subgraphs from graph and piped through a graph neural network that combines convolutional component and recurrent neural networks to perform the information cascade prediction of cascade graphs. CasCN learning the latent representation of cascade graph through the temporal features and structure features.

\subsection{ Evaluation Metric}
\label{sec_metric}
We follow previous studies and use standard measurement strategies – MSLE (mean square log2-transformed error) in our experiments \cite{Li2016DeepCas}, \cite{cao2017deephawkes}, \cite{chen2019information}. Note that the smaller the MSLE, the higher the prediction accuracy. Specifically, MSLE is defined as:
\begin{equation}\label{MSLE}
MSLE= \frac{1}{N}\sum\limits_{i=1}^{N}{(log_{2}{\widetilde{G}_{i}^{\Delta t}}}-log_{2}{G_{i}^{\Delta t})^2}
\end{equation}
In the equation, $N$ is the total number of citation networks. $\widetilde{G}_{i}^{\Delta t}$ is the predicted growth of citation counts, and $G_{i}^{\Delta t})^2$ is the real growth of citation counts.

\subsection{Parameters Setting}
\label{sec_set}
\label{Parameters_Setting}
In order to make it easy for scholars to repeat our experiments, we will list the detailed parameters of each method in this section.

Firstly, for Node2Vec, we follow the work in ~\cite{grover2016node2vec}, i.e., parameters p and q are selected from $\{0.25, 0.50, 1, 2, 4\}$, the length of
walk is chosen from $\{10, 25, 50, 75, 100\}$, and the number of walks per node varies from $\{5, 10, 15, 20\}$.

For DeepCas, DeepHawkes and Topo-LSTM, we use the default settings in DeepCas, where the embedding dimensionality of nodes is 50, the hidden layer of each GRU has 32 units and the hidden dimensions of the two-layer MLP are 32 and 16, respectively. The learning rate for user embeddings is ${5*10{-}4}$ and the learning rate for other variables is ${5*10{-}3}$. The batch size for each iteration is 32. The training process will stop after 1000 iterations. For CasCN, in addition, we choose the support k = 2 of GCN and calculate the maximum eigenvalue $\lambda$max of cascade Laplacian. Other settings are basically consistent with DeepCas.

As for our DeepCCP, there is one important parameters: time interval. Different datasets have different time interval. Therefore, we find the optimal parameters for different datasets through experiments. The specific parameters are shown in the Table ~\ref{PARADENN}.

\begin{table}[]
\caption{Parameter Configuration of DeepCCP.}\label{PARADENN}
\centering
\begin{tabular}{ccc}
\hline
             &Time Interval& Degree Sequence Length \\ \hline
APS-5years   & 6  &[114,84,71,57,41,39,33,41,30,14,12,7,4,3,2]\\
APS-7years   & 7  &[134,85,72,58,53,48,45,52,35,26,20,11,7,7,7,3,2,6,2]\\
APS-9years   & 7  &[138,103,79,60,61,69,54,56,39,31,33,31,19,12,10,6,5,13,9,6,6,3]\\
Hep-PH-3years  & 6 &[109,75,51,51,49,52,43,52,45,39,28,24,20,21,12,9,4,1,1]\\
Hep-PH-4years & 8  &[123,86,71,63,52,56,52,76,72,71,52,40,28,30,20,20,19,8,5,2]\\
Hep-PH-5years & 7  &[128,98,81,70,52,56,52,76,72,71,52,40,28,30,20,20,19,8,5,5,1]\\ \hline
\end{tabular}
\end{table}

\subsection{Performance}
\label{sec_performance}
Our experiments on HEP-PH dataset and APS dataset verify the high performance of DeepCCP. In this section, we show the comparison results of DeepCCP and other baseline methods in detail to illustrate the effectiveness and stability of DeepCCP in the paper citation count prediction.

\begin{figure}
  \centering
  \includegraphics[width=\textwidth]{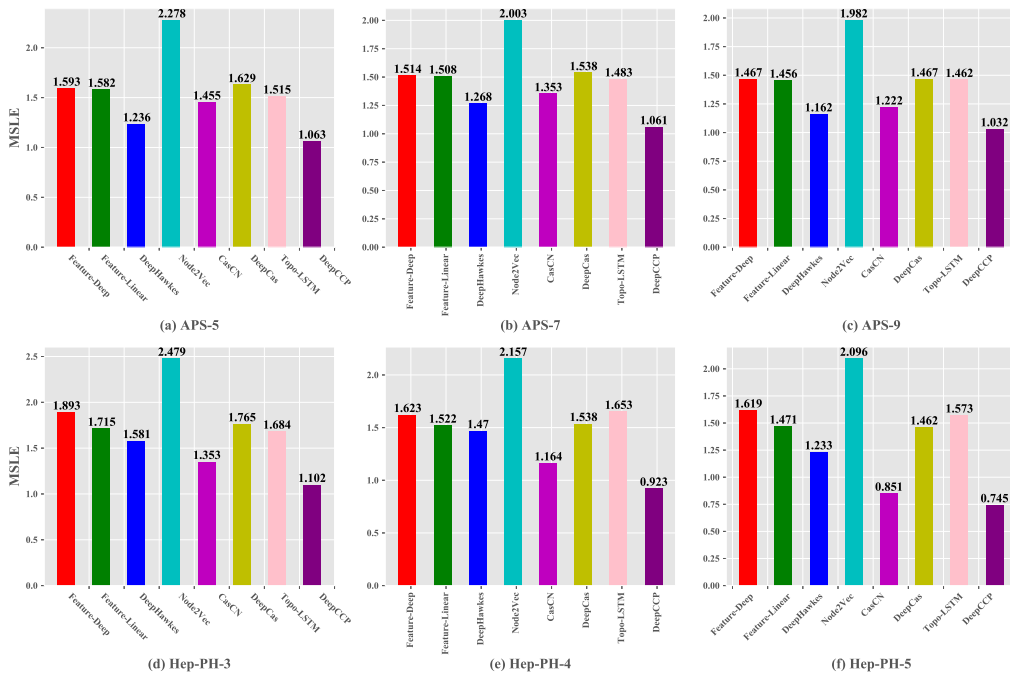}
  \caption{DeepCCP vs. Baselines.}\label{results}
\end{figure}

\noindent\textbf{DeepCCP vs. Baselines} For the feature-based baseline method: we can conclude that the performance of Feature-Linear and Feature-Deep in the Hep-PH and APS data sets is relatively stable and at a moderate level, almost just better than Node2Vec. Moreover, to our surprise, we thought that feeding the features of the citation network into the MLP module would result in better performance. However, experimental results show that Feature-Linear is almost always better than Feature-Deep. This tells us that deep learning methods do not always work. This experimental result also tells us that the feature-based method can also achieve good results in the field of citation counts prediction, of course, the premise is that you can find suitable features.

As shown in the Figure \ref{results}, for the network embedding method: the Node2Vec method has the worst performance on the two data sets, which also shows that the method of sampling sequences from the network based on random walk cannot effectively extract the structure of the network, hence, simply use this method for predicting citation counts is inappropriate.

For the method based on deep learning, DeepCas also utilizes the random walk mechanism of the Node2Vec method, but adds the attention mechanism and GRU, so it performs better than Node2Vec. Although DeepCas is superior to the feature-deep method and Node2Vec in both data sets, DeepCas is inferior to the feature-linear method in APS data set, which once again shows that the method of deep learning is not necessarily superior to the feature-based method; Topo-LSTM utilize a topological Long Short-Term Memory module to predict citation counts, but this method still cannot extract enough information from citation networks, so there is no outstanding performance in both datasets. As for DeepHawkes and CasCN, These two methods use the structure and temporal information of the citation network at the same time, while improving the interpretability, they also have good performance on the two data sets.

As for DeepCCP, it can be seen from the experimental results of the Figure \ref{results} that the best effect has been achieved in both datasets, and all the baseline methods have been defeated, which demonstrate that our method significantly outperforms the state-of-the-art methods for paper citation counts prediction. For example, it achieves excellent prediction results with MSLE = 0.923 when observing for 4 years in Hep-PH and MSLE = 1.061 when observing for 5 years in APS, respectively. It reduces the prediction error by 20.7\% and 16.3\% comparing to the second best CasCN and DeepHawkes, respectively.

\subsection{Discussing the Degree Distribution Sequences}
\label{sec_Discussing}
We believe that DeepCCP can capture some of the underlying structure of the citation network. In this section, we will try to verify our statement.

Our approach is to feed the vector calculated by the formula directly into an MLP module. The output of this MLP is some structural features of the citation network, such as: edges, max path length, average path length, leaf node numbers and average degree, and then calculate the MSE between the outputs of MLP and the true features. If the MSE is very small, close to 0, then it can be shown that the vector can fit this feature well, and vice versa. In other words, we use vectors to fit the structural features of the corresponding citation network, and use the fitting results to determine whether the vectors contain structural information.

The experimental results are shown in the table \ref{inputdata_fit}. We can see that the MSE of each feature in the table is close to 0. Therefore, our vector does contain rich structural information, which is evidence that DeepCCP can capture the citation network structure. And in DeepCCP, it has a stronger network structure, can capture richer structural information, and integrates the temporal information of the citation network, so it can obtain  the state-of-the-art results in the application of citation counts predicting.

\begin{table}[]
\caption{Analysis of Degree Sequences and Citation Network Structural Features.}\label{inputdata_fit}
\centering
\begin{tabular}{lccccc}
\hline
                          & edges                & max\_path            & ave\_path            & leaves               & ave\_degree          \\ \hline
\multicolumn{1}{c}{Hep-PH} & 0.0004               & 0.0331               & 0.0022               & 0.0004               & 0.0005               \\
\multicolumn{1}{c}{APS}   & 0.0005               & 0.0402               & 0.0113               & 0.0007               & 0.0003               \\ \hline
                          & \multicolumn{1}{l}{} & \multicolumn{1}{l}{} & \multicolumn{1}{l}{} & \multicolumn{1}{l}{} & \multicolumn{1}{l}{}
\end{tabular}
\end{table}

\subsection{Analysis of Parameter Sensitivity}
\label{sec_Sensi}
In order to verify the robustness of DeepCCP, in this chapter, we conducted a sensitivity analysis experiment. In DeepCCP, there is only one key parameter: time interval. Specifically, in the sensitivity analysis experiment, we choose Hep-PH 4years and APS 7years data sets, and observe the changes in MSLE by changing the value of the time interval.

\begin{figure}
  \centering
  \includegraphics[width=\textwidth]{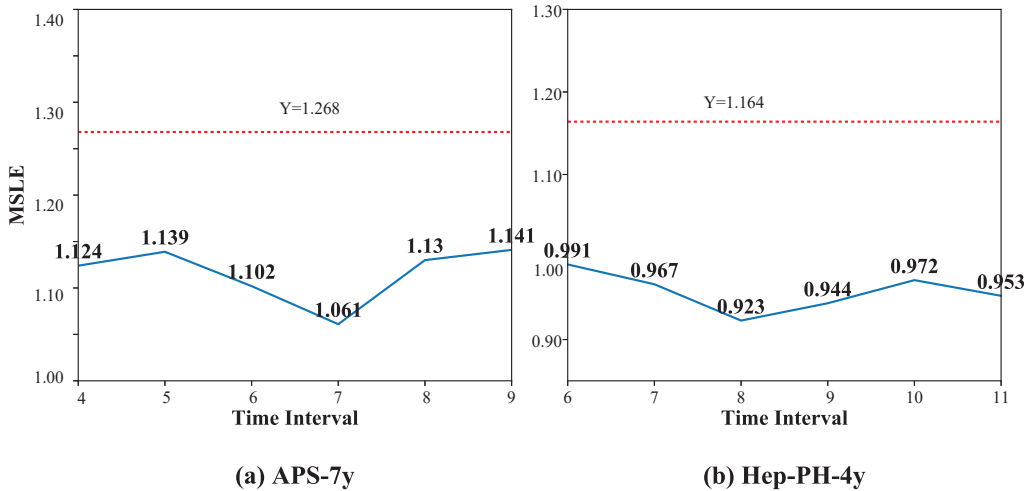}
  \caption{ Analysis of Time Interval Sensitivity. The red horizontal line represents the optimal value in the baseline methods.}\label{TIME_SENSE}
\end{figure}

As shown in the Figure \ref{TIME_SENSE}, as the time interval changes, MSLE is constantly changing. On the two data sets of APS and Hep-PH, the changes of MSLE are relatively stable and always higher than other optimal baseline methods. Hence, we believe that DeepCCP is relatively robust.

\section{Conclusion And Future Work}
\label{sec_conclusion}
In this paper, we propose an end-to-end deep learning framework for paper citation counts prediction: DeepCCP, which combines the information cascade effect to dynamically model the information cascade citation network formed in the early stage of the paper, and then predict the paper Citations. DeepCCP proposes a degree sequence citation network representation method, and uses 4 modules of pre-embedding layer, time Delay module, GRU module and convolutional neural network to deeply extract the structure and temporal information of the paper citation network. Our comprehensive experiments on two types of 6 data sets have verified the superiority of DeepCCP in the field of citation counts predicting compared with other baseline methods, which reflects the state-of-the-art method.

As for future work, on the one hand, our DeepCCP only uses the structural information of the paper citation network and the temporal information generated by the paper citation. However, there is still a lot of information that is not used, such as author characteristics, paper content, etc., these information It also has a great influence on paper citation prediction, so in future work we will consider embedding this information in deep learning neural networks; on the other hand, we think that deep learning networks can be moderately deepened or new modules can be added , such as auto-encoder, could strengthen the ability to extract information, and then obtain better performance.
%



\bibliographystyle{elsarticle-num}
\bibliography{mybibliography}



\end{document}